\documentclass[a4paper,11pt]{article}

\usepackage{jinstpub} 
\usepackage{subfig}

\usepackage{placeins}
\usepackage[ngerman,english,nameinlink]{cleveref}
\crefname{chapter}{Chapter}{Chapters}
\crefname{section}{Sec.}{Sec.}
\crefname{subsection}{Sec.}{Sec.}
\crefname{figure}{Fig.}{Fig.}
\crefname{table}{Tab.}{Tab.}
\crefname{appendix}{Appendix}{Appendix}
\crefname{equation}{Eq.}{Eq.}

\usepackage{lineno}
\linenumbers

\title{\boldmath Characterization of the TruSense S310 Laser Range System for Contact-less Measurement of Liquid Levels in Large-Volume Neutrino Detectors}

\author[a,b,c,1]{H. Th. J. Steiger,\note{Corresponding author.}}
\author[a,b]{E. Theisen,}
\author[c]{L. Oberauer,}
\author[a,b]{O. Pilarczyk,}
\author[a,b]{M. Wurm,}

\affiliation[a]{Cluster of Excellence PRISMA$^+$ \\Staudingerweg 9, 55128 Mainz, Germany}
\affiliation[b]{Institute of Physics, Johannes Gutenberg University Mainz \\ Staudingerweg 7, 55128 Mainz, Germany}
\affiliation[c]{Physik-Department E15, Technische Universität München \\ James-Franck-Straße 1, 85748 Garching, Germany}

\emailAdd{hsteiger@uni-mainz.de}

\abstract{Neutrino experiments often use large volumes of water, organic scintillators or noble liquids as active detection material. Due to the large hydrostatic and buoyancy forces involved, precise knowledge of the liquid levels inside the detector tank are mandatory. Here we present the main characteristics of the TruSense S310 Laser Range System. Level measurements can be performed without direct contact to the liquid and through a gas-proof acrylic window, thus preserving the strict radiopurity and chemical requirements of the target liquid. We report the results of a suit of laboratory experiments for short-term precision tests ($\pm$5\,mm) and long-term stability studies. Moreover, we demonstrate that the infrared laser can be used while standard bi-alkali PMTs are operational. We discuss the mechanical layout and integration of the system in the OSIRIS pre-detector that will monitor the radiopurity of the liquid scintillator for the large-volume neutrino experiment JUNO.}


\arxivnumber{1234.56789} 

\begin{document}
\nolinenumbers
\maketitle
\flushbottom

\section{Introduction}
Most neutrino experiments use large volumes filled with liquid scintillator (LS), noble liquids or water for neutrino detection. In many cases, the detectors feature several concentric liquid tanks or otherwise complicated geometries. This makes for challenging filling procedures that require a precise monitoring of filling level heights inside the integrated tanks. Similar monitoring also is required for the vessels of the attached liquid handling system (LHS) and in major storage tanks. Even small differences in filling levels (typically of the order of 20 cm) between these subsystems may lead to the appearance of strong hydrostatic or buoyancy forces that can cause fatal damage to the detector vessels. For the usage in JUNO and OSIRIS a precision of $\pm$5 cm and a stability of the measurement values of $\pm$2.5 cm over time can be considered as minimum design requirement. Moreover, the filling level of the detectors should be sampled with a repetition rate $\geq$0.01 Hz.  

In handling those liquids, high demands are set in terms of radioactive purity of target liquids and detector materials since they are a central aspect to achieve the ultra-low backgrounds required for the detection of neutrinos. This sets the additional requirement that the determination of the filling level heights must not compromise the quality of the detector liquids used for the experiment. Contact with a contaminated sensor float or intrusion of ambient air into the system has to be ruled out in order to maintain strict radiopurity standards.

\begin{figure}[t!]\centering
	\includegraphics[width=0.4\textwidth]{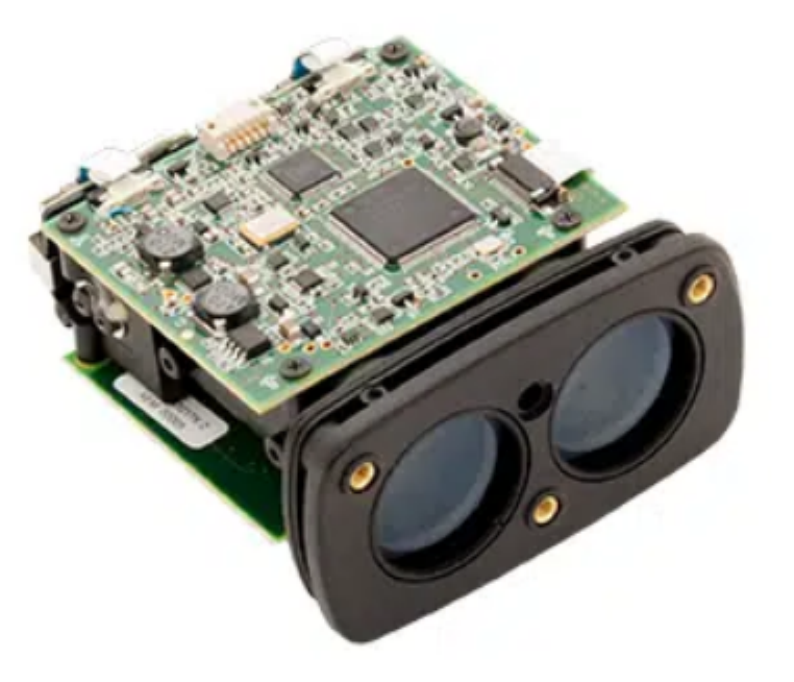}
\quad
	\includegraphics[width=0.53\textwidth]{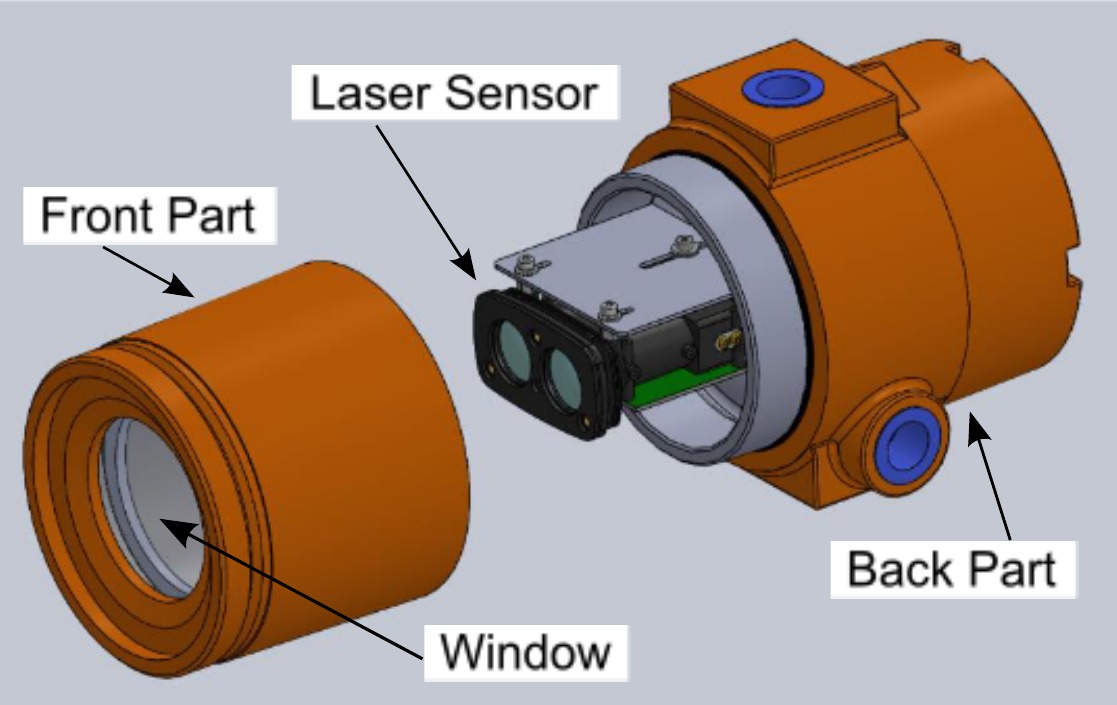}
\caption{\textbf{Left:} Commercial photo of the TruSense S310 in the OEM version with diffusor lense taken from the website of the supplying company \cite{LaserTec_web}. \textbf{Right:} 3D-layout of the TruSense S310 mounted inside its enclosure. In this illustration the enclosure front is screwed off the back part allowing a vision of the sensor. The scheme is taken from the user's manual \cite{TruSense}.}
\label{fig:TruSense}
\end{figure}

The present paper investigates the TruSense S310 laser range system for use in such an environment. The TruSense S310 system is produced by Laser Technology Inc.~(depicted in \cref{fig:TruSense}). It is capable of measuring the distance to a liquid surface while avoiding direct contact between the monitoring system and the target medium. The maximum operating distance is 50\,m. The liquid level height is determined based on a time-of-flight (TOF) measurement performed by means of a pulsed IR (infra-red) laser ($\lambda$ =905 nm \cite{TruSense}). The system directly exploits the laser pulse reflections at the liquid surface and thus enables the determination of the level without the additional application of a guided float. Furthermore, the measurement can be performed through a gas-proof acrylic window ensuring that the target is still sealed against environmental influences. This new method has several advantages compared to the laser-based level measurement system of the Double Chooz Near Detector ~\cite{Franke}, since a contamination of the detector liquids in consequence of direct contact with a float or due to the entrance of oxygen or radioactive gases is excluded. Thus, the application of the IR laser range system preserves the strict radiopurity and chemical requirements on the detector material. 

The paper assesses all the steps necessary to implement the TruSense S310 in a large-volume liquid detector. In \cref{sec:characterisation} we present the main characteristics and components of the IR laser sensor as well as the software interface for operation. As a working example, \cref{sec:integration} describes the mechanical integration of the TruSense S310 system in the level-measurement system of the OSIRIS  (Online Scintillator Internal Radioactivity Investigation System) pre-detector \cite{osiris}. The 9-by-9-meter OSIRIS setup will be used as an online monitor for the scintillator radiopurity in the JUNO (Jiangmen Underground Neutrino Observator) scintillator filling chain \cite{juno-ppnp}. \cref{sec:tests} presents a series of laboratory experiments dedicated to evaluate the performance of the TruSense S310: short-term precision and long-term stability of the measurement and finally the simultaneous operation of the IR laser with standard bi-alkali photomultiplier tubes (PMTs).
\cref{sec:DC} reports the results of a test deployment of the TruSense S310 in the realistic detector environment of the Double-Chooz Near Detector setup \cite{DC}. This \textit{in-situ} trial run of the system became possible during the draining operations for the Near Detector, permitting a comparison of the capabilities of the laser-range system with a standard hydrostatic pressure sensor.

\section{General Properties of the TruSense S310 Laser Sensor}
\label{sec:characterisation}
The TruSense S310 features a number of characteristics that make it particularly attractive for use in a large-volume neutrino detector:
\begin{itemize}
    \item The level monitoring is performed by optical means which enables a contact-less filling level measurement preserving the radiopurity of the inserted detector liquids.
    \item No guided float in direct contact to the detector liquid is required, as the distance measurement is performed by the direct reflection of laser light from a transparent liquid surface.
    \item The system allows the measurement through a protective window which seals off the detector volume against the entrance of oxygen or radioactive gases.
    \item The system allows for constant monitoring of the insertion and draining process and does not require an interruption of the respective procedures to operate properly.
    \item Simultaneously running bi-alkali PMTs are not harmed by potentially occurring reflexes of the infra-red laser pulses emitted by the system. 
\end{itemize}
The contact-less measurement of the laser-range system is in contrast to that of other sensor types, most notably hydrostatic pressure sensors and capacitive contact sensors. Both have to be in direct contact with the detector liquid, setting high demands on radiopurity levels and material compatibility of the sensor surfaces. Even compared to other laser-range systems used on earlier occasions, e.g.~the M10L/100 \cite{MEL} from MEL-Mikroelektronik deployed in the Double Chooz experiment, the TruSense S310 offers the advantage that neither a float in contact with the liquid nor the deactivation of the PMTs during a measurement is required \cite{Franke}.


The following paragraphs describe the primary components and characteristics of the TruSense S310 and briefly discusses the software interface for the operation of the system.

\subsection{Sensor Characteristics}
The TruSense S310 infrared laser sensor has been designed for non-contact fluid level measurement of highly reflective and turbulent surfaces in distances in the range from $\sim$~0.5 m to 50 m to the sensor front window~\cite{TruSense}. The system can be used for continuous monitoring of rising and lowering liquid surfaces and does not require an interruption of the filling procedure to provide an accurate measurement\footnote{This is different from hydrostatic pressure sensors when mounted on a filling pipe in which case Bernoulli's law will lower the static pressure.}. The distance to a target surface is determined by measuring the return time from the emission of an infrared laser pulse to the arrival of the corresponding target reflex and comparing it to the speed of light (TOF technique). The ranging laser is intensity class 1 and emits pulses at a wavelength of 905~nm. The frequency of these pulses can be customized in the range from 1\,Hz to 14\,Hz, which corresponds to the number of distance measurements performed per second. In addition, the orientation of the beam path can be reviewed by a visible class~2 alignment laser with a wavelength of 650\,nm~\cite{TruSense}.

The TruSense S310 supports the application of three target modes using either the first reflex~(DF~mode), the strongest reflex~(DS~mode) or the last reflex~(DL~mode) caused by the target surface. However, in most test measurements we observed only minor differences between these three measurement options (\cref{sec:tests}).

The system mainly used for testing in the laboratory frame is the OEM (original equipment manufacturer) version without a housing. \cref{fig:TruSense} displays the laser system as well as the optional housing made of glass fiber reinforced plastic. It protects the sensor interior against an accumulation of dust. To prevent any damage to the electronics and optics of the system in consequence of external impacts or fire, the sensor is mounted inside a safety proved cylindrical enclosure (orange) with a diameter of $\sim$~127~mm, a length of $\sim$~254~mm and a weight of $\sim$~3.8~kg~\cite{TruSense}. The enclosure front and back parts are covered with lids that also prevent the entrance of dust particles and that can be unscrewed to gain access to the interior. The front lid is equipped with a viewing window in order to enable the operation of the laser sensor inside the enclosure.

The TruSense S310 requires an input power of 12 - 24 VDC which can be provided by a default power socket using the supplied Power COM Cable (PCC). The data sheet states a power draw of 1.8\,W for the sensor in operation and of 0.48\,W when in stand-by mode~\cite{TruSense}.

\subsection {Beam Diameter and Diffuser Lens}
The free aperture of the measurement laser features a width of $\sim$23\,mm. Together with the stated beam divergence of $\sim$3\,mrad, this yields a beam diameter of the $\sim$173\,mm at the maximum measurement range of 50\,m~\cite{TruSense}. In case of OSIRIS (\cref{sec:integration}), the beam path is led through a pipe into the interior of the acrylic vessel. In this setup the inner pipe diameter provides a constraint to the maximally feasible beam spread, since reflections at the inner pipe wall have to be excluded to ensure a proper measurement (\cref{sec:tests}). For level height measurements of turbulent liquid surfaces the manufacturer recommends the application of an additional diffuser lens attachable to the front of the laser sensor. The use of the diffusor lens results in an increased spread of the laser beam diameter. For this reason, the system will be operated without this lens in OSIRIS.

\subsection{Sensor Communication and Software Interface}
For data transmission between user and sensor the TruSense S310 offers the serial RS-232 ASC II communication pathway. This standard method can be applied to read out the sensor ouput or to modify the sensor settings. In addition, a standardized SDI-12 serial interface can be applied~\cite{TruSense}, but has never been used in the scope of the present work. Beside the realization of the power supply, the PCC also establishes the connection between laser sensor and a serial I/O device. The PCC is supplied with a DB9 pin serial connector which requires the application of RS-232-to-USB adapter, if a connection to a standard USB port is desired.

\begin{figure}[t!]\centering
	\includegraphics[width=0.85\textwidth]{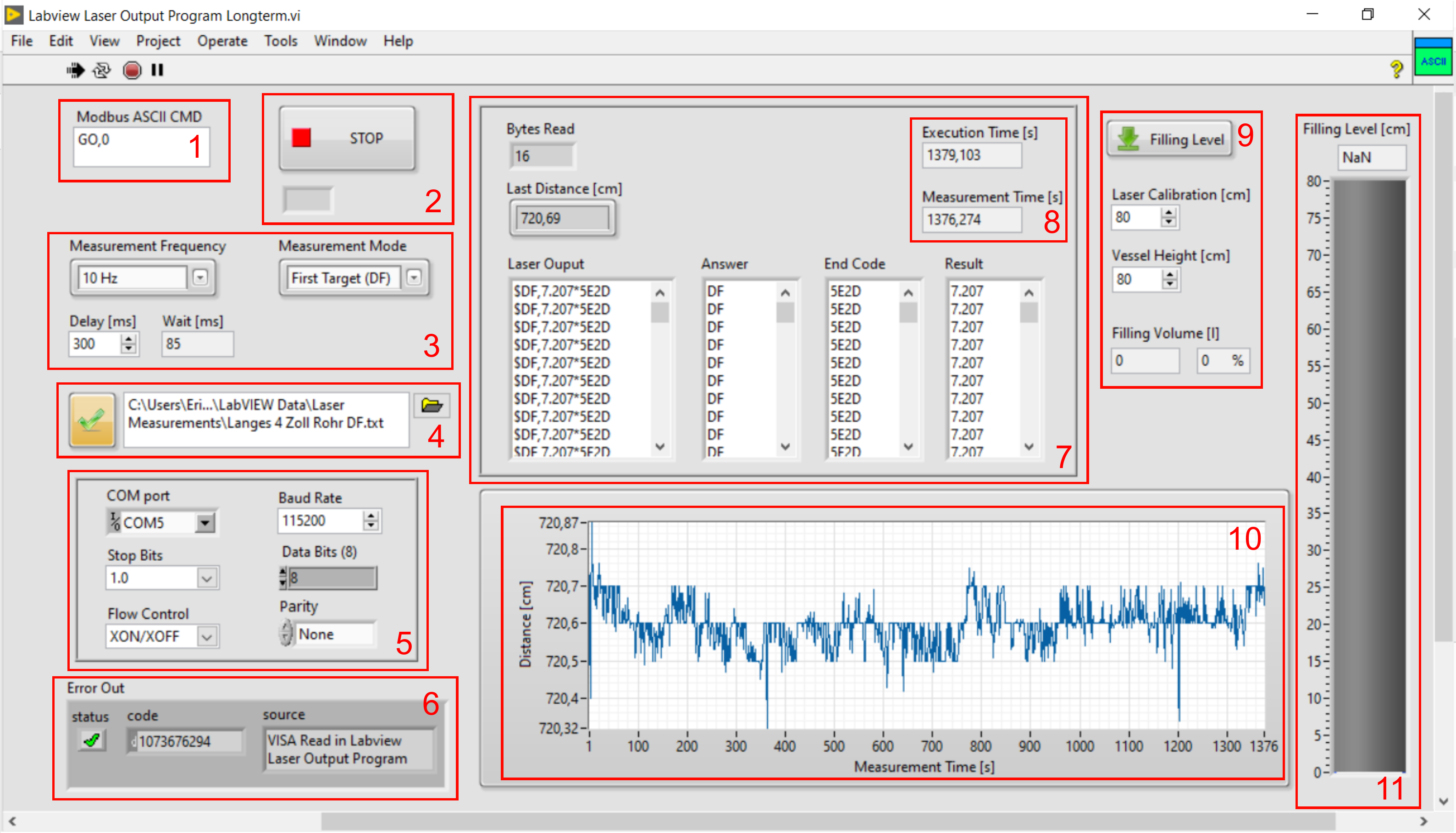}
\caption{Front panel screenshot of the LabVIEW software interface developed and used for the communication with the TruSense S310 laser sensor. Numeration: \;1)\nobreakspace Command input \;2)\nobreakspace Stop button with laser status display \;3)\nobreakspace Adjustments of measurement frequency, target mode, time delay and wait period \;4)\nobreakspace Data storage to file \;5)\nobreakspace Communication settings \;6)\nobreakspace Error output \;7)\nobreakspace String display of measurement results \;8)\nobreakspace Display of execution and measurement time \;9)\nobreakspace Button to switch to the automatic conversion of the distances measured to filling level heights. For this purpose, the entries of the laser mounting height and vessel height are required. \;10)\nobreakspace Graphical filling level display \;11)\nobreakspace Automatic plot of the distances recorded against the measurement time}
\label{fig:software}
\end{figure}

As communication and read out software the Laser Technology Inc. recommends a terminal emulation program, such as Tera Term \cite{TruSense, TeraTerm}. However, within the scope of this present work a user interface under the application of the visual programming environment LabVIEW~(Laboratory Virtual Instrument Engineering Workbench~\cite{LabVIEW}) developed by the National Instruments Corporation~(NI) has been created. We benefited from the collection of sample codes offered at the NI community support website which provided a basic virtual environment (VI) to read out and write ASC II register values via a serial RS-232 modbus \cite{examplecode}. The VI offers a command line to enter RS-232 ASC II serial commands and additional input boxes to manage the communication settings (for further information, please refer to \cite{TruSense}). After the duration of an adjustable read delay, the response of the laser sensor is displayed as a string output along with the number of bits read. In addition, the sample code offers a display for error messages. Through the consecutive integration of further features to the sample VI, a comprehensive LabVIEW user interface has been developed enabling a convenient and more sophisticated usage of the TruSense S310. The interface includes selection boxes to choose the measurement frequency and the target mode for the upcoming measurement. Depending on the frequency selected, the program automatically calculates the means of the distances measured per second and displays the measurement time. Results, means and the associated measurement times are stored in lists that can be written to a text file and saved at a costumizable location on the hard disk of the connected device. For a distance measurement performed with 10 Hz over a period of 15 min the created text-file requires a disk space of $\sim$ 275 kB which corresponds to $\sim$ 10 GB per year for a measurement in continuous operation (exceeding by far the required repetition rate of $\geq$0.01 Hz). Due to this small amount of data, the easy direct readability of text files (.txt) is preferred over more common compressed binary files. Furthermore, the LabVIEW interface is equipped with a graphical display showing a live plot of the means determined from the collected data. In addition, we added a feature which automatically converts the distances measured to filling level heights. \cref{fig:software} presents the layout of the LabVIEW user interface as used for the laboratory test measurements described in \cref{sec:tests}. 

\begin{figure}[t!]\centering
	\includegraphics[width=0.7\textwidth]{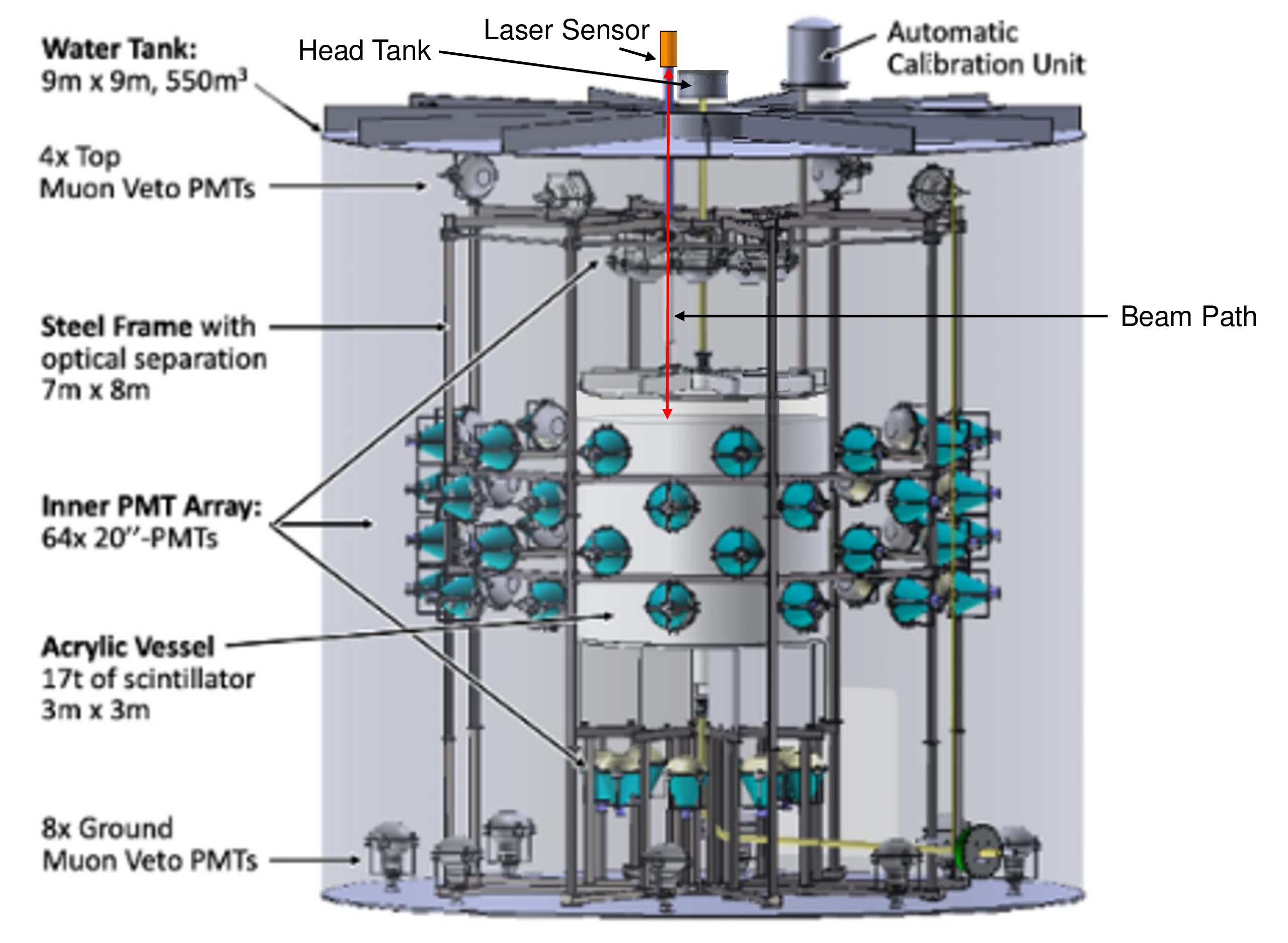}
\caption{Schematic overview of the OSIRIS pre-detector layout with a labeling of the central component parts including the laser sensor installation \cite{osiris}.}
\label{fig:OSIRIS}
\end{figure}

\section{Integration in the OSIRIS Detector Setup}
\label{sec:integration}
The design of the OSIRIS pre-detector is schematically shown in \cref{fig:OSIRIS}. OSIRIS has been designed to monitor the radiopurity of the JUNO liquid scintillator (LS) \cite{osiris}. These measurements will be performed both during the commissioning of the JUNO scintillator mixing and purification systems as well as during the extended filling phase of the JUNO main detector. Six months will be required to fill 20,000 tons of LAB-based LS into the JUNO main detector \cite{juno-ppnp}. 

The OSIRIS setup holds 20\,m$^{3}$ (18\,t) of LS to directly observe and determine the decay rates of radioactive trace elements dissolved in the liquid. The LS is contained in a cylindrical acrylic tank of 3\,m height and diameter that is positioned in the center of the detector setup. For shielding from external gamma-ray backgrounds, the acrylic vessel is surrounded by a 9-by-9 meter cylindrical steel tank that holds a water volume of 550 m$^3$ \cite{osiris}. The scintillation light from radioactive decays in the LS is detected by an array of 64 Hamamatsu 20-inch-PMTs submerged in the water volume, arranged on a stainless steel frame at a distance of 1.3\,m from the outer surface of the acrylic vessel. Twelve additional PMTs mounted to the top and floor of the water tank detect Cherenkov photons created by the passage of cosmic muons through the water buffer, providing an anti-coincidence veto for cosmic muons. A more detailed description of the setup can be found in \cite{osiris}.

\begin{figure}[t!]\centering
	\includegraphics[width=1\textwidth]{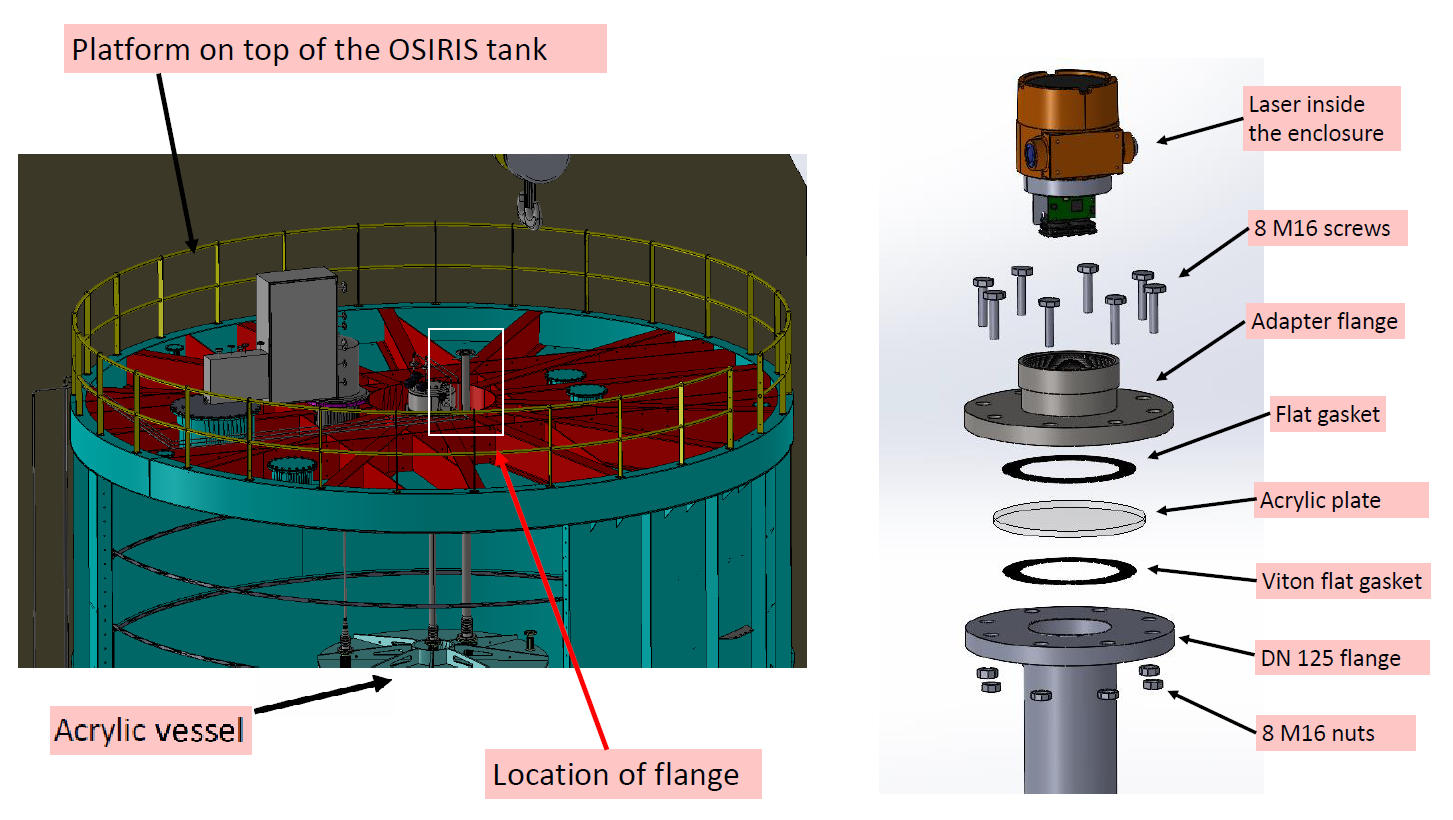}
\caption{\textbf{Left:} 3D-Layout of OSIRIS showing the designated location of the laser sensor installation at the roof of the OSIRIS water tank. \textbf{Right:} Schematical illustration of the components parts required for laser sensor installation indicating their intended assembly.}
\label{fig:installation}
\end{figure}

The TruSense S310 system is used in this setup primarily to monitor the LS liquid level during detector filling and to monitor the stability of the filling level during regular operation. 
The left panel of \cref{fig:installation} shows a schematic view of the integration of the laser system into the OSIRIS setup. The laser sensor is mounted above the top lid of the water tank, 50\,cm above the nominal level of the LS during detector operation at the end of a 4-inch stainless steel pipe. This $\sim$4 m long pipe provides a direct line of sight into the inner volume of the acrylic vessel (bottom $\sim$ 7\,m below the laser). This allows to perform measurements during detector filling as soon as the liquid level reaches the lower lid of the acrylic vessel. The connection between laser sensor and pipe is realized by the interface depicted in the right panel of \cref{fig:installation}. The pipe finishes in a flat-face DN125 flange (EN1092-1, PN16), while the laser sensor is mechanically mounted to a corresponding counterpart flange. This so-called adapter flange has been specifically designed for this application to adapt the dimensions of the DN125 flange to the 4-inch NPT screw thread that fits the back part of the laser-sensor enclosure provided by the manufacturer. The two flanges form a sandwich around an acrylic viewing window and thus provide an air-tight seal of the interface, separating the inner pipe volume and thus acrylic vessel from the ambient air. Gas in- and outlets mounted to the sides of the pipe permit to maintain a nitrogen atmosphere above the liquid inside the pipe, preventing the contact of scintillator with atmospheric oxygen and radioactive gases. In accordance to the EN1092-1 norm, the flanges are connected by a set of eight M16 screws (12\,Nm torque) and sealed with FKM flat gaskets on both sides of the window pane.

\section{Test Measurements and Lab Setup}
\label{sec:tests}
To assure that the TruSense S310 fulfills all the requirements for the save operation of the OSIRIS detector (\cref{sec:integration}), we have performed a suit of test measurements. These investigations reviewed the precision of the laser sensor and the stability of the sensor output in long-term measurements. Furthermore, we ran several tests to validate the mechanical integration of the laser system in the OSIRIS geometry, in particular the diameter of the laser pipe and the distance as well as the thickness of the protective window just below the laser sensor. In this section, we report on the experimental setups used for these tests and discuss the main results.

\begin{figure}[t!]\centering
\includegraphics[width=0.65\textwidth]{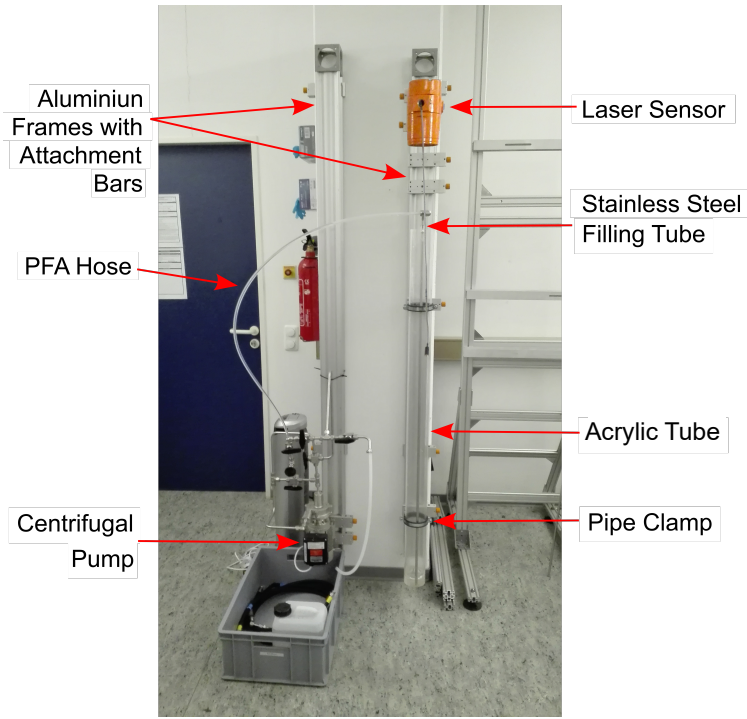}
\caption{View of the laboratory setup built up to investigate the laser sensor functionality. The acrylic tube with an inner diameter of 90 mm was connected via a PFA hose and a stainless steel filling tube to a centrifugal pump used to fill and drain the tube with liquids like water or LAB. The laser sensor was mounted on an aluminium frame directly above the acrylic tube to measure the liquid level inside the tube.}
\label{fig:labsetup}
\end{figure}

\subsection{Laser Sensor Precision}
\begin{figure}[t!]\centering
	\includegraphics[width=1\textwidth]{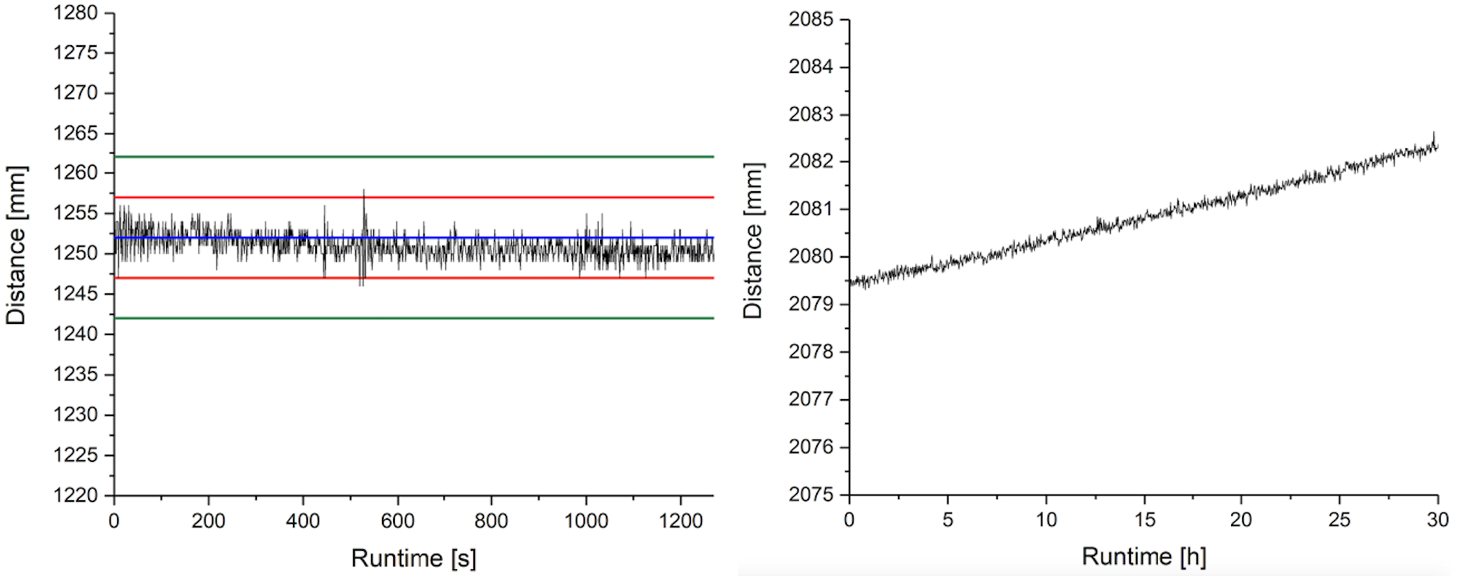}
\caption{\textbf{Left:} Distance values to a liquid surface with a sampling rate of 1~Hz. The mean value of the raw sensor readings from the first 200 s is plotted as blue solid line, while the red lines indicating the edges of a $\pm$5~mm band around the mean. The green lines represent the maximum fluctuation of the sensor readings of $\pm$10~mm specified by Laser Technology Inc. \textbf{Right:} Long-term measurement of the distance to a water level in the acrylic pipe. The data points in the plot are the mean values obtained from 100~sampling points for the distance recorded at a rate of 1~Hz. Clearly visible is an increase in the distance corresponding to a falling water level in the pipe due to evaporation of water during the 30~h of measurement duration.}
\label{fig:LabTest}
\end{figure}

The precision of a distance measurement performed by the TruSense S310 is stated by the manufacturer as $\pm 10$~mm~\cite{TruSense}. With the help of a dedicated experimental setup we reviewed this statement in a series of measurements. For this purpose, an acrylic tube was filled with a highly transparent liquid (e.g. water or LAB) and subsequently the distance to the liquid surface was measured with the laser sensor mounted at an aluminium frame right above the tube. The precise adjustment of the laser central into the acrylic pipe and orthogonal to the liquid level was ensured by means of a plumb bob and the sensors internal visible alignment laser (for details see \cite{TruSense}). A photo of the experimental setup is presented by \cref{fig:labsetup}.\\
\\
\textbf{Short-term measurements:}\\
In a first test measurement, the sensor was operated with a repetition rate of 1~Hz for $\sim$20~min. \cref{fig:LabTest}~left shows the measured values for the most intense laser reflex (DS mode). The mean value of the measured data points (from the first 200 sampling points) gives a distance of (1252$\pm$0.1)~mm, which is in excellent agreement with the  cross-check by means of a precision steel ruler. All measurement values lie within a $\pm$10~mm band around the mean value, which corresponds to the maximum reading fluctuation specified by Laser Technology Inc. Except for a few single data points, all values are within a band of $\pm$5~mm around the mean. The readings for the first (DF mode) and last (DL mode) reflex correspond to those shown here within $\pm$2~mm.\\ 
\\
\textbf{Long-term measurement:}\\
For the usage in neutrino detectors slow changes in the liquid level need to be resolved precisely. Therefore, drifts in the sensor response are very critical for the determination of small increases / decreases in distance to a liquid surface. For this investigation, the slow change in liquid level was realized by the slow evaporation of water from the setup's acrylic pipe. Before the measurement was started, the water level was marked on the pipe by means of a felt-tipped pen. Subsequently, data was taken with a sampling rate of 1~Hz for 30~h. For the entire measurement the temperature was kept stable within $\pm$0.2~°C by the laboratory's air conditioning system. In order to make the fine changes in the level clearly visible, the mean values calculated from 100 sampling points were plotted in \cref{fig:LabTest}~right. These method exploits the close to normal distribution of the raw data points around their corresponding mean. Starting at a distance to the water surface of $\sim$2079.5~mm the readings slightly increased during the 30 h to $\sim$2082.4~mm. This discrepancy was confirmed comparing the marked water level on the tube before and after the measurement using a vernier caliper. The ability of the sensor to resolve these small and slow fluctuations of the liquid level exceeds the requirements for OSIRIS. Averaging a selectable number of sampling points for precision increase was therefore implemented in the sensor software interface.   

\FloatBarrier

\subsection{Simultaneous Operation of Laser System and Bi-Alkali PMTs}

\begin{figure}[t!]
    \centering
    \includegraphics[width=1.00\textwidth]{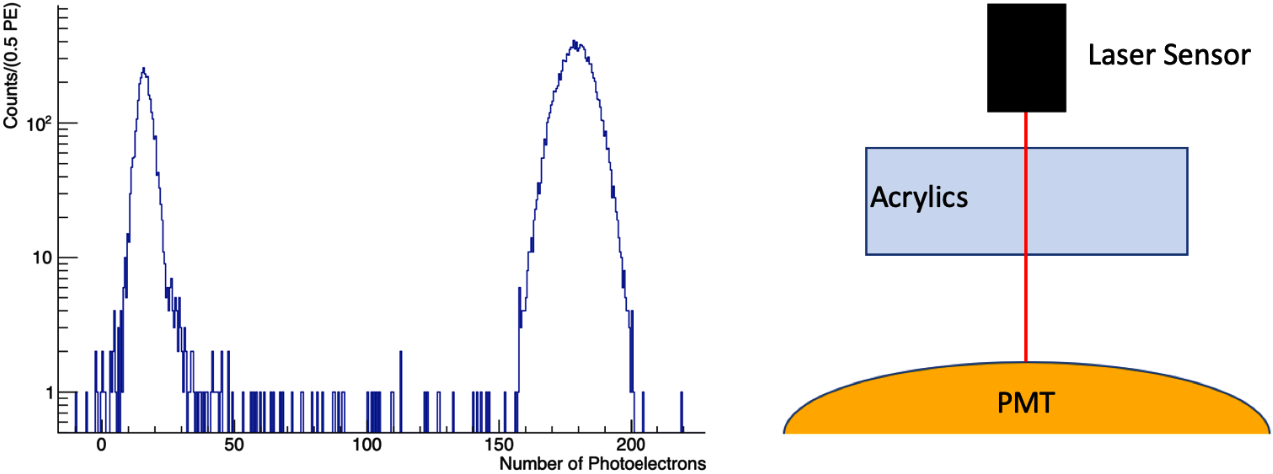}
    \caption{\textbf{Left:} Charge spectrum of the PMT obtained from the IR laser pulses shining through 12 cm of highly transparent acrylic glass. The peak at $\sim$180 p.e. corresponds to radiation directly hitting the photo cathode, while the left peak is caused by IR photons undergoing previous reflections on the interfaces between acrylics and air. \textbf{Right:} Shown is the arrangement used for the PMT irradiation test. The laser sensor was placed at a distance of $\sim$1 m to the central part of the PMT photo cathode. Within the laser beam a 12 cm thick block of highly transparent acrylics was placed to emulate the impact of JUNO’s central detector sphere. Note, in JUNO the IR light will be further attenuated by $\sim$2 m of water between the PMTs and the acrylic sphere.}
    \label{fig:PMT_Test}
\end{figure}

To ensure that the laser sensor can be operated during the operation of the OSIRIS pre-detector and potentially also the JUNO main detector without posing a hazard to the PMTs, a Hamamatsu H12860HQE PMT (type to be used in both detectors, see also \cite{PmtPaper}) has been irradiated with light from the IR laser sensor during operation. The laser irradiated the photo cathode through 12 cm of highly transparent acrylic glass, emulating the effect, corresponding to the thickness of the acrylic sphere of the JUNO Central Detector \cite{juno-ppnp}. The sensor and the acrylic were placed in a distance of $\sim$1 m to the PMT, the beam aiming directly at the PMT photocathode. For the irradiation test, the laser was operated without the diffuser lenses resulting in a smaller beam divergence of 3 mrad and with the highest possible measurement repetition rate of 14 Hz. The arrangement, which can be considered as a worst case scenario for the usage in JUNO and OSIRIS, is shown on the right side of Figure \cref{fig:PMT_Test}. The spectrum on the left illustrates that the laser pulse directly hitting the photocathode causes signals corresponding to about 160 to 200 photo electrons. IR radiation reflected at the interfaces between acrylic glass and air, and reaching the photocathode with a short time delay caused a second peak in the spectrum of about 17 photo electrons.

While the corresponding signals are large compared to regular scintillation events (typically < 10 p.e.) and would thus interfere with regular detector operation, the signal amplitudes are not dangerous, posing no risk of destruction of the PMTs. In the OSIRIS geometry, such a direct line of sight will be realized only for a single PMT directly below the laser sensor. Stray light from the reflection on the LS surface might affect further PMTs. This means that operation of the PMTs could be continued during active operation of the laser system, provided that light signals observed in coincidence with the laser pulses are excluded from the physics data set. The introduced detector deadtime can be considered to be negligible (< 1min per day). It should be mentioned here, that in JUNO the IR light will be further attenuated by $\sim$2 m of water between the PMTs and the acrylic sphere.

\FloatBarrier

\section{Tests in the Double Chooz Near Detector}
\label{sec:DC}
\begin{figure}[t!]
    \centering
    \includegraphics[width=0.38\textwidth]{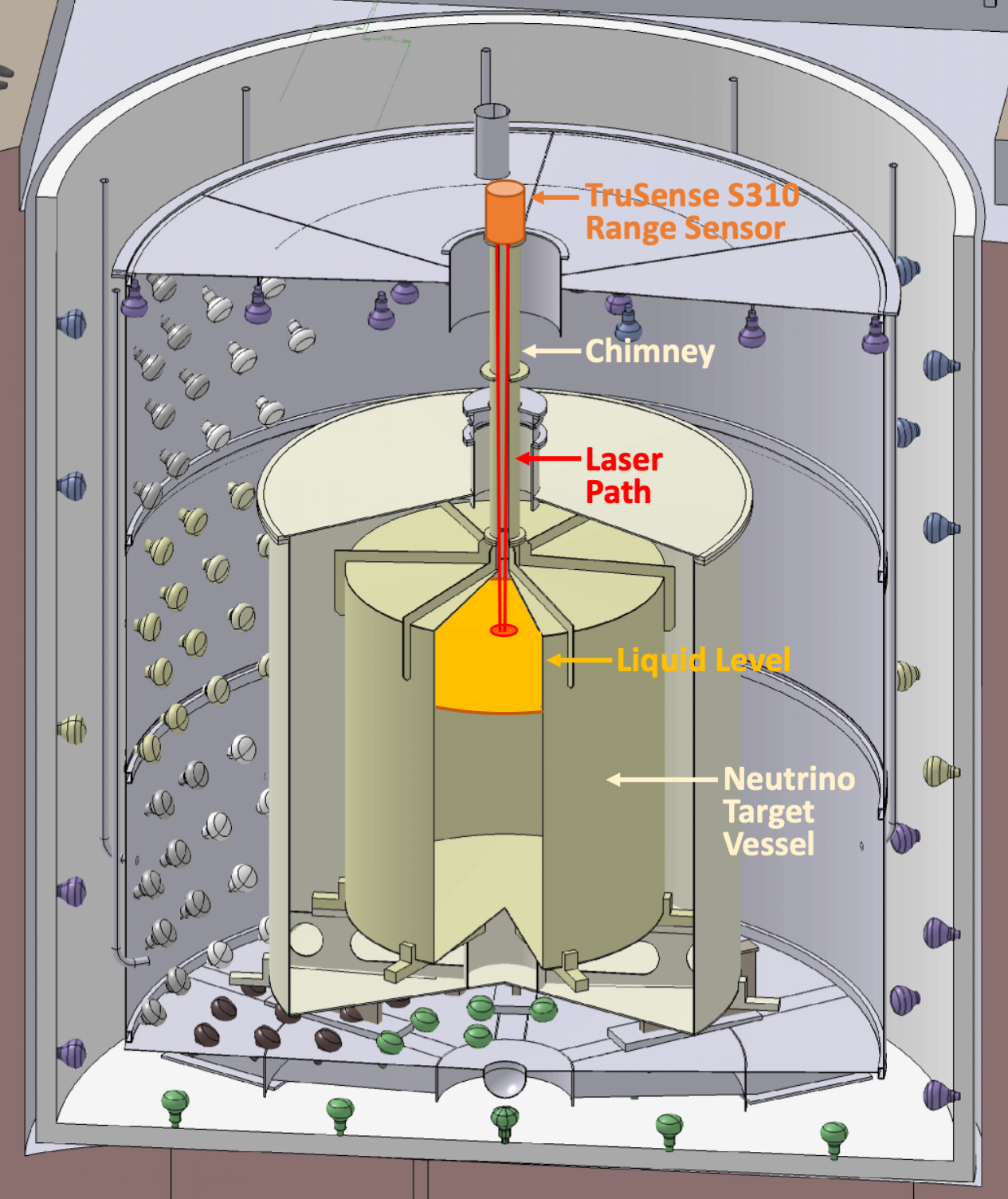}~~~~
    \includegraphics[width=0.55\textwidth]{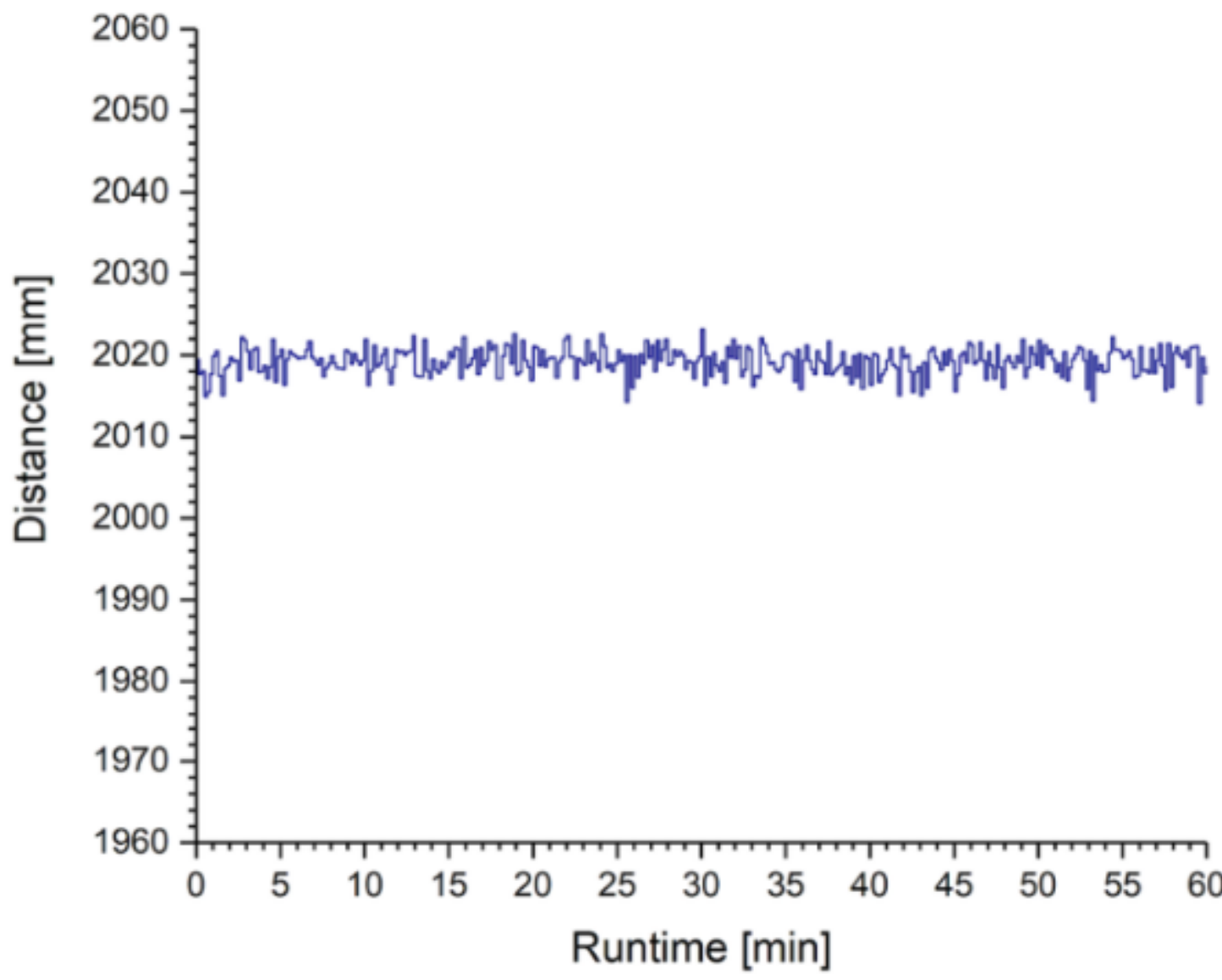}
\caption{\textbf{Left:} Schematic view of the TruSense S310 Laser Range Sensor mounted onto the chimney of the Double Chooz Near Detector. The (static) liquid level inside the Neutrino Target volume was measured through a vertical connection pipe as indicated by the read beam path. \textbf{Right:} Time series of the distance measurement to the static scintillator surface. A sampling rate of 1\,Hz is applied for a total run time of an hour. Individual data points average over 10 seconds.}
    \label{fig:DC_Test}
\end{figure}

The dismantling of the Near Detector of the Double Chooz reactor neutrino experiment~\cite{DC} offered the opportunity to perform a sensor test under the operation conditions of a full-scale neutrino experiment. The TruSense S310 was mounted on top of an acrylic pipe (Detector Chimney) reaching down into the central liquid scintillator volume, the Neutrino Target (see \cite{DC} for details). The laser range system was mounted in the position of the mechanical level measurement system (Tamago, Proservo NMS5) shown in the left panel of \cref{fig:DC_Test} (taken~from~\cite{DC}), using a customized mechanical fixation for the laser sensor. The liquid level inside the Detector Chimney was measured through the opened ball valve for a period of an hour. The measured values for the distance to the surface of the highly transparent LS inside the detector are depicted in the right panel of \cref{fig:DC_Test}. The obtained distance to the LS of (2019.2~$\pm$~0.1(stat.)) mm is in excellent agreement with the expectation (200~$\pm$~1.5) cm calculated from the live readings of the hydrostatic pressure sensors on the bottom of the Neutrino Target vessel. The stability of the sensor readings well within 1~cm of fluctuation tolerance is remarkable. 

\FloatBarrier


\section{Conclusions}

Precise monitoring of the liquid filling levels inside the tanks and vessels of large-scale neutrino experiments is of great importance to prevent damage to the detector integrity caused by unintentional occurrence of hydrostatic pressure differences and buoyancy forces. A contact-less and sealed laser range system like the TruSense S310 can at the same time maintain the high radiopurity standards for the detector liquids. The series of tests presented in this paper has been sufficient to assure that the laser range system can be used and mechanically integrated in the OSIRIS detector setup. In a short-term measurement, the sensor was operated with a repetition rate of 1~Hz for $\sim$20~min. All measurement values were found within a $\pm$10~mm band around the mean value, which corresponds to the maximum reading fluctuation specified by Laser Technology Inc. Except for a few single data points, all values are within a band of $\pm$5~mm around the mean. The long-term stability of the measurement results has been deduced from a 30 h measurement. Since the deviation of the measured values can be fully explained by the evaporation of the target liquid, a sub-mm stability of the averaged measured values from 100 sampling points can be concluded.\\
Furthermore, we have developed an integration scheme for the TruSense S310 in the OSIRIS detector setup that can be easily adapted to a wide range of detector sizes and geometries. In a further setup we verified that a simultaneous operation of the laser system and standard bi-alkali PMTs is feasible. Finally, the successful deployment and operation of the TruSense S310 atop the Double Chooz Near Detector represents a first practical test of the system in a realistic detector environment. We conclude that the TruSense S310 is potentially applicable to a wide range of detector geometries and liquids in future neutrino experiments.

\acknowledgments

We would like to thank Rainer Othegraven and Karl-Heinz Geib (JGU Mainz, ETAP) for their substantial technical support in the design and installation of the laboratory setups and OSIRIS detector interface, and Tobias Sterr (University of Tübingen, AG Lachenmaier) for his support for the PMT irradiation tests. This work was supported by funds of the DFG Research Unit FOR2319 ``JUNO'', the Cluster of Excellence PRISMA$^+$ and the Maier Leibnitz Laboratorium (MLL).




\end{document}